\documentclass[12pt]{article}
\usepackage[left=1.8cm,right=1.8cm, top=2cm, bottom=2cm]{geometry}
\usepackage{graphics,graphicx,color,amsfonts, bm, amsfonts, amssymb, todonotes}
\usepackage[authoryear]{natbib}
\usepackage[width=.95\linewidth,justification=centering]{caption}

\usepackage{algorithmic}
\algsetup{linenosize=\small}
\usepackage{algorithm}

\newcommand{\comment}[1]{}

\newcommand{\bmx}[0]{{\bm x}}
\newcommand{\bmy}[0]{{\bm y}}
\newcommand{\bmlam}[0]{{\bm \lambda}}

\title{Exact Bayesian Inference for the Bingham Distribution}
\author{Christopher J. Fallaize \& Theodore Kypraios\footnote{Author for correspondence: {\tt theodore.kypraios@nottingham.ac.uk}} \\ \small{School of Mathematical Sciences, University of Nottingham, Nottingham, NG7 2RD, UK}}
\date{\today}

\begin{document}
\maketitle

\abstract{This paper is concerned with making Bayesian inference from data that are assumed to be drawn from a Bingham distribution. A barrier to the Bayesian approach is the parameter-dependent normalising constant of the Bingham distribution, which, even when it can be evaluated or accurately approximated, would have to be calculated at each iteration of an MCMC scheme, thereby greatly increasing the computational burden. We propose a method which enables exact (in Monte Carlo sense) Bayesian inference for the unknown parameters of the Bingham distribution by completely avoiding the need to evaluate this constant. We apply the method to simulated and real data, and illustrate that it is simpler to implement, faster, and performs better than an alternative algorithm that has recently been proposed in the literature}.

\section{Introduction} \label{sec:intro}

Observations that inherit a direction occur in many scientific disciplines \citep[see, for example,][]{MarJup00}. For example, directional data arise naturally in the biomedical field for protein structure \citep{Boom_etal08}, cell--cycle \citep{RuedFerPed09} and circadian clock experiments \citep{ Lev_etal02}; see also the references in \citet{EhlerGal11}. A distribution that has proved useful as a model for spherical data which arise as unsigned directions is the Bingham distribution \citep{Bin74, MarJup00}.

% introduce to the reader the Bingham Distribution
The Bingham distribution can be constructed by conditioning a zero-mean multivariate Normal (MVN) distribution to lie on the sphere $\mathcal{S}^{q-1}$ of unit radius in $\mathbb{R}^q$. In particular, for a given matrix $A$ of dimension $q \times q$, the density with respect to the uniform measure on $\mathcal{S}^{q-1}$ is given by % $\mbox{d}_{\mathcal{S}^{q}} (\bmx)$
\begin{equation}
f(\bmx; A) = \frac{\exp{(-\bmx^T A \bmx)}} {c(A)}, \quad \bmx^T \bmx = 1 \mbox{ and } \bmx \in \mathbb{R}^{q},
\label{eq:density_Bingham}
\end{equation}
where $c(A)$ is the corresponding normalising constant.

% inference for the Bingham
Having observed some directional data, interest then lies in inference for the matrix $A$ in (\ref{eq:density_Bingham}). The likelihood of the observed data given the parameters can easily be written down and at first glance it appears that maximum likelihood inference for $A$ is straightforward. However, inferring the matrix $A$ is rather challenging. That is due to the fact that the likelihood of the observed data given the matrix $A$ involves the parameter-dependent normalising constant $c(A)$ which, in the general case, is not available in closed form. Therefore this poses significant challenges to undertake statistical inference involving the Bingham distribution either in a frequentist or Bayesian setting.

% what people have done and why our approach is worth doing and/or is better than the others?
Although a maximum likelihood estimator for $A$ can be derived by iterative techniques which are based on being able to approximate $c(A)$ \citep[see, for example,][]{Ken87, KumWood05, KumWood07, TomKum13}, very little attention has been drawn in the literature concerning estimation of $A$ within a Bayesian framework. \citet{Wal13} considered Bayesian inference for the Bingham distribution which removes the need to compute the normalising constant, using a (more general) method that was developed earlier \citep{Wal11} and cleverly gets around the intractable nature of the normalising constant. However, it requires the introduction of several latent variables and a Reversible-Jump Markov Chain Monte Carlo (RJMCMC) sampling scheme. 
  
The main contribution of this paper is to show how one can draw Bayesian inference for the  matrix $A$, by exploiting the recent developments in Bayesian computation for distributions with doubly intractable normalising constants \citep{Mol_etal06, MurGhahMac06}. The main advantage of our approach is that it does not require any numerical approximation to $c(A)$ and hence, enables {\em exact} (in the Monte Carlo sense) Bayesian inference for $A$. Our method relies on being able to simulate {\em exact} samples from the Bingham distribution which can be done by employing an efficient rejection sampling algorithm proposed by \citet{Ken12}.  

% The plan of the paper
The rest of the paper is structured as follows. In Section \ref{sec:rejection_general} we introduce the family of Angular Central Gaussian distributions and illustrate how such distributions serve as efficient proposal densities to sample from the Bingham distribution. In Section \ref{sec:bayesian_inf} we describe our proposed algorithm while in Section \ref{sec:application} we illustrate our method both using simulated and real directional data from earthquakes in New Zealand. In Section \ref{sec:concl} we discuss the computational aspects of our method as well as directions for future research.

\section{Rejection Sampling} \label{sec:rejection_general}

\subsection{Preliminaries}
Rejection sampling \citep{Rip87} is a method for drawing independent samples from a distribution with probability density function $f(x) = f^*(x)/Z_f$ assuming that we can evaluate $f^*(x)$ for any value $x$, but may not necessarily know $Z_f$. Suppose that there exists another distribution, with probability density function $g(x) = g^*(x)/Z_g$, often termed an {\em envelope density}, from which we can easily draw independent samples and can evaluate $g^*(x)$ at any value $x$. We further assume that there exists a constant $M^*$ for which $M^* g^*(x) \geq f^*(x)$ $\forall x$. We can then draw samples from $f(x)$ as follows:

{\tt
\begin{enumerate}
 \item Draw a candidate value $y$ from $g(x)$ and $u$ from $U(0,1)$;
 \item if $u \leq \frac{f^*(y)}{M^* g^*(y)}$ accept $y$; otherwise reject $y$ and go to step 1.
\end{enumerate}
}

The set of accepted points provides a sample from the target density $f(x)$. It can be shown that the number of trials until a candidate is accepted has a geometric distribution with mean $M$, where 
\begin{equation}
M = \sup_{x \in \mathcal{R}}\left\{\frac{f(x)}{g(x)} \right\} < \infty. \label{eq:constant_M}
\end{equation}
Therefore, the algorithm will work efficiently provided that  $M$ is small or, in other words, the probability of acceptance $(1/M)$ is large. Moreover, it is important to note that it is not necessary to know the normalising constants $Z_f$ and $Z_g$ to implement the algorithm; the only requirement is being able to draw from the envelope density $g(x)$ and knowledge of $M^*$ (rather than $M$).

\subsection{The Angular Central Gaussian Distribution}
The family of the angular central Gaussian(ACG) distributions is an alternative to the family of the Bingham distributions for modelling antipodal symmetric directional data \citep{Tyl87}. An angular central Gaussian distribution on the $(q-1)-$dimensional sphere  $\mathcal{S}^{q-1}$ can be obtained by projecting a multivariate Gaussian distribution in $\mathbb{R}^q$, $q \geq 2$, with mean zero onto $\mathcal{S}^{q-1}$ with radius one. In other words, if the vector ${\bm y}$ has a multivariate Normal distribution in $\mathbb{R}^q$ with mean ${\bm 0}$ and variance covariance matrix $\Psi$, then the vector $\bmx = {\bm y}/|| {\bm y}||$ follows an ACG distribution on $\mathcal{S}^{q-1}$ with $q \times q$ symmetric positive$-$definite parameter matrix $\Psi$ \citep{MarJup00}. The probability density function of the ACG distribution with respect to the surface measure on $\mathcal{S}^{q-1}$ is given by 
\begin{equation}
g(\bmx; \Psi) = w_q^{-1}|\Psi|^{-1/2} \left(\bmx^T \Psi^{-1} \bmx \right)^{-q/2} = c_{\mathrm{ACG}}(\Psi) g^*(\bmx; \Psi)
\end{equation}
where the constant $w_q = 2\pi^{q/2}/\Gamma(q/2)$ represents the surface area on $\mathcal{S}^{q-1}$. Denote by $c_{\mathrm{ACG}}(\Psi) =  w_q^{-1}|\Psi|^{-1/2} $ the normalising constant where $\Psi$ is a $q \times q$ symmetric positive$-$definite matrix. 

\subsection{Rejection Sampling for the Bingham Distribution} \label{sec:rejection}
\citet{Ken12} have demonstrated that one can draw samples from the Bingham distribution using the ACG distribution as an envelope density within a rejection sampling framework. In particular, the following algorithm can be used to simulate a value from the Bingham distribution with parameter matrix $A$: \vspace{0.2cm}

{\tt
\begin{enumerate}
 \item Set $\Psi^{-1} = I_q + \frac{2}{b} A$ and $M^* \ge \sup_{\bm{x} }\left\{\frac{f*(x)}{g*(x)} \right\}; $ 
 \item draw $u$ from $U(0,1)$ and a candidate value ${\bm y}$ from the ACG distribution on \linebreak the sphere with parameter matrix $\Psi$;
 \item if $u < \frac{f^*({\bm y}; A)}{M^{*} g^*({\bm y}; \Psi)}$ accept ${\bm y}$; otherwise reject ${\bm y}$ and go to Step 1.
\end{enumerate}
}\vspace{0.2cm}

Here, $f^*(\bm{y};A) = \exp(-\bm{y}^T A \bm{y}) $, and $g^*(\bm{y}; \Psi) = (\bm{y}^T \Psi^{-1} \bm{y})^{-\frac{q}{2}}$, the unnormalized Bingham and ACG densities respectively, and $b < q$ is a tuning constant. We found that setting $b=1$ as a default works well in many situations, but an optimal value can be found numerically by maximising the acceptance probability $1/M$ \citep[see, for example,][]{Gan12}.

%In particular, it was shown that for some  constant $b<q$ then upper bound $M$ in (\ref{eq:constant_M}) is equal to 
%\begin{eqnarray}
% M(q, b; A, \Psi) & = & \frac{c_{\mbox{Bing}}(A)}{c_{\mathrm{ACG}}(\Psi} \left(\frac{q}{b}\right)^{q/2} \exp\left\{-\frac{1}{2}(q- b)\right\} \nonumber \\
%    & = & \frac{c_{\mbox{Bing}}(A)}{c_{\mathrm{ACG}}(\Psi)} M^* \label{eq:constant_M_bing}
%\end{eqnarray}
%Equation (\ref{eq:constant_M_bing}) involves a tuning constant $b$. Although it is possible to derive the optimal value for $b$ when $q = 2$, it is not as easy to do so in practice when $q \geq 3$ \citep{Gan12}. Having obtained $M^*$ and setting $\Psi^{-1} = I_q + \frac{2}{b} A$ we can employ the following rejection sampling algorithm: 

\section{Bayesian Inference} \label{sec:bayesian_inf}
\subsection{Preliminaries}
Consider the probability density function of the Bingham distribution as given in (\ref{eq:density_Bingham}). If $A = V \Lambda V^T$ is the Singular Value Decomposition of $A$ where $V$ is orthogonal and $\Lambda = \mbox{diag}(\lambda_1, \ldots, \lambda_q)$, then it can be shown that if $\bmx$ is drawn from a distribution with probability density function $f(\bmx; A)$, the corresponding random vector ${\bm y} = X^T V$ is drawn from a distribution with density $f(\bmx; \Lambda)$ \citep[see, for example,][]{KumWal06, KumWood07}. Therefore, without loss of generality, we assume that $A = \Lambda = \mbox{diag}(\lambda_1, \ldots, \lambda_q)$. Moreover, to ensure identifiability, we assume that $\lambda_1 \geq \lambda_2 \geq \ldots \lambda_q = 0$ \citep{Ken87}. Therefore, the probability density function becomes
\begin{equation}
 f(\bmx; \Lambda) = \frac{\exp \left\{-\sum_{i=1}^{q-1}\lambda_i x_i^2\right\}}{c(\Lambda)} \label{eq:density_Bingham_lambda}
\end{equation}
with respect to a uniform measure on the sphere and 
$$
c(\Lambda) = \int_{\bmx \in S^{q-1}} \exp\left\{ -\sum_{i=1}^{q-1}\lambda_i x_i^2 \right\} \mbox{ d} S^{q-1}(\bmx).
$$

%\subsection{Likelihood}
Suppose $({\bmx_1, \bmx_2, \ldots, \bmx_n})$ is a sample of unit vectors in $\mathcal{S}^{q-1}$ from the Bingham distribution with density (\ref{eq:density_Bingham_lambda}). Then the likelihood function is given by 
\begin{equation}
 L(\Lambda) = \frac{1}{c(\Lambda)^{n}} \exp\left\{- \sum_{i=1}^{q-1} \lambda_i \sum_{j=1}^{n} \left(x_j^i\right)^2 \right\} = \frac{1}{c(\Lambda)^{n}} \exp\left\{-n\sum_{i=1}^{q-1} \lambda_i  \tau_i \right\}, \label{eq:lik_Bingham}
\end{equation}
where $\tau_i = \frac{1}{n}\sum_{i=1}^{n}\left(x_j^i\right)^2$. The data can therefore be summarised by $(n, \tau_1, \ldots, \tau_{q-1})$, and $(\tau_1, \ldots, \tau_{q-1})$ are sufficient statistics for $(\lambda_1, \ldots, \lambda_{q-1})$. 

\subsection{Bayesian Inference}
We are interested in drawing Bayesian inference for the matrix $\Lambda$, or equivalently, for ${\bm \lambda} = (\lambda_1, \ldots, \lambda_{q-1})$. The likelihood function in (\ref{eq:lik_Bingham}) reveals that the normalising constant $c(\Lambda)$ plays a crucial role. The fact that there does not exist a closed form expression for $c(\Lambda)$ makes Bayesian inference for $\Lambda$ very challenging. 

For example, if we assign independent Exponential prior distributions to the elements of ${\bm \lambda}$ with rate $\mu_i$ (i.e. mean $1/\mu_i$) subject to the constraint that $\lambda_1 \geq \lambda_2 \geq \ldots \geq \lambda_{q-1}$ then the density of the posterior distribution of $\Lambda$ up to proportionality given the data is as follows:
\begin{eqnarray}
 \pi({\bm \lambda}|\bmx_1, \ldots, \bmx_n) & \propto & L(\Lambda) \prod_{i=1}^{q-1} \exp\{-\lambda_i \mu_i\} \cdot  {\bm 1}\left(\lambda_1 \geq \lambda_2 \geq \ldots \geq \lambda_{q-1} \right) \nonumber \\
 & = & \frac{1}{c(\Lambda)^{n}} \exp \left\{ - \sum_{i=1}^{q-1} \lambda_i (n \tau_i + \mu_i) \right\} \cdot  {\bm 1}\left(\lambda_1 \geq \lambda_2 \geq \ldots \geq \lambda_{q-1} \right).  \label{eq:post_dens_Bingham}
\end{eqnarray}

Consider the following Metropolis-Hastings algorithm which aims to draw samples from $\pi({\bm \lambda}|\bmx_1, \ldots, \bmx_n)$:

{\tt
\begin{enumerate}
 \item Suppose that the current state of the chain is ${\bm \lambda}^{\mbox{cur}}$;
 \item Update ${\bm \lambda}$ using, for example, a random walk Metropolis step by \newline proposing ${\bm \lambda}^{\mbox{can}} \sim N_{q-1}\left({\bm \lambda}^{\mbox{cur}}, \Sigma\right)$; 
 \item Repeat steps 1-2.
\end{enumerate}
}
Note that $N_{q-1} \left({\bm m}, S \right)$ denotes  the density of a multivariate Normal distribution with mean vector ${\bm m}$ and variance-covariance matrix $S$. Step 2 of the above algorithm requires the evaluation of the ratio $\pi\left({\bm \lambda}^{\mbox{can}}|{\bmx_1, \ldots, \bmx_n}\right)/\pi\left({\bm \lambda}^{\mbox{cur}}|{\bmx_1, \ldots, \bmx_n}\right)$, which in turn involves evaluation of the ratio $c(\Lambda^{\mbox{can}})/c(\Lambda^{\mbox{cur}})$. Therefore, implementing the above algorithm requires an approximation of the normalising constant. In principle, one can employ one of the proposed methods in the literature which are based either on asymptotic expansions \citep{Ken87}, saddlepoint approximations \citep{KumWood05}  or holonomic gradient methods \citep{TomKum13}. Although such an approach is feasible, in practice, it can be very computationally costly since the normalising constant would have to be approximated at every single MCMC iteration. Furthermore, despite how accurate 
these approximations may be, the stationary distribution of such an MCMC algorithm won't be the distribution of interest $\pi({\bm \lambda}|\bmx_1, \ldots, \bmx_n)$, but an approximation to it. 

\subsubsection{An Exchange Algorithm} \label{sec:exchange}

The main contribution of this paper is to demonstrate that recent developments in Markov Chain Monte Carlo algorithms for the so-called doubly intractable distributions enable drawing exact Bayesian inference for the Bingham distribution without having to resort to any kind of approximations. 

\citet{Mol_etal06} proposed an auxiliary variable MCMC algorithm to sample from doubly intractable distributions by introducing a cleverly chosen variable in to the Metropolis-Hastings (M-H) algorithm such that the normalising constants cancel in the M-H ratio. In order for their proposed algorithm to have good mixing and convergence properties, one should have access to some sort of typical value of the parameter of interest, for example a pseudo-likelihood estimator. A simpler version that avoids having to specify such an appropriate auxiliary variable was proposed in \citet{MurGhahMac06}. Although both approaches rely on being able to simulate realisations from the Bingham distribution (see Section \ref{sec:rejection}), we choose to adapt to our context the approach presented in \citet{MurGhahMac06} because it is simple and easy to implement, since a value of the parameter of interest does not need to be specified.  

Consider augmenting the observed data with auxiliary data $\bmy$, so that the corresponding augmented posterior density becomes

\begin{equation}
\pi(\bmlam,\bmy,|\bmx) \propto \pi(\bmx|\bmlam) \pi(\bmlam)  \pi(\bmy|\bmlam), \label{eq:augm_post}
\end{equation}
where $\pi(\bmy |\bmlam)$ is the same distribution as the original distribution on which the data $\bmx$ is defined (i.e. in the present case, the Bingham distribution). Proposal values for updating the parameter $\bmlam$ are drawn from a proposal density $h(\cdot | \bmlam)$, although in general this density does not have to depend on the variables $\bmlam$. For example, random walk proposals centred at  $\bmlam$ or independence sampler proposals could be used.

%Consider the augmented distribution with the following probability density 
%\begin{equation}
%\pi(\bmlam^{\prime},\bmx^{\prime}, \bmlam|\bmx) \propto \pi(\bmx|\bmlam) \pi(\bmlam) h(\bmlam^{\prime}|\bmlam) \pi(\bmx^{\prime}|\bmlam^{\prime}) \label{eq:augm_post}
%\end{equation}
%where $\pi(\bmx^{\prime}|\bmlam^{\prime})$ is the same distribution as the original distribution on which the data $\bmx$ is defined (i.e. the Bingham). The distribution distribution $h(\bmx^{\prime}|\bmx)$ could be any arbitrary distribution for the augmented variables $\bmlam^{\prime}$ which may, although it does not have to, depend on the variables $\bmlam$, for example, a random walk distribution centered at  $\bmlam$. It is easy to see that the marginal distribution of $\bmlam$ in (\ref{eq:augm_post}) is the posterior distribution of interest.

Now consider the following algorithm: \vspace{0.2in}
{\tt
\begin{enumerate}
 \item Draw $\bmlam^{\prime} \sim h(\cdot | \bmlam)$;
 \item Draw $\bmy \sim \pi(\cdot|\bmlam^{\prime})$;
 \item Propose the exchange move from $\bmlam$ to $\bmlam^{\prime}$ with probability 
$$
\min\left(1, \frac{f^*(\bmx|\bmlam^{\prime}) \pi(\bmlam^{\prime}) h(\bmlam|\bmlam^{\prime}) f^*(\bmy|\bmlam)}
{f^*(\bmx|\bmlam) \pi(\bmlam) h(\bmlam^{\prime}|\bmlam) f^*(\bmy|\bmlam^{\prime})}  \times \frac{c(\Lambda) c(\Lambda^{\prime})}{c(\Lambda) c(\Lambda^{\prime})}\right),
$$
\end{enumerate}
}\vspace{0.2in}
%Now consider the following algorithm: \vspace{0.2in}
%{\tt
%\begin{enumerate}
% \item Draw $\bmlam^{\prime} \sim h(\cdot | \bmlam)$
% \item Draw $\bmx^{\prime} \sim \pi(\cdot|\bmlam^{\prime})$
% \item Propose the exchange move from $\bmlam$ to $\bmlam^{\prime}$ with probability 
%$$
%\min\left(1, \frac{g(\bmx^{\prime}|\bmlam) \pi(\bmlam^{\prime}) h(\bmlam|\bmlam^{\prime}) g(\bmx|\bmlam^{\prime})}
%{g(\bmx|\bmlam) \pi(\bmlam) h(\bmlam^{\prime}|\bmlam) g(\bmx^{\prime}|\bmlam^{\prime})}  \times \frac{c(\Lambda) c(\Lambda^{\prime})}{c(\Lambda) c(\Lambda^{\prime})}\right)
%$$
%\end{enumerate}
%}\vspace{0.2in}
where $f^*(\bm{x};A) = \exp(-\bm{x}^T A \bm{x}) $ is the unnormalized Bingham density as previously. This scheme targets the posterior distribution of interest (the marginal distribution of $\bmlam$ in (\ref{eq:augm_post})), but most importantly, note that all intractable normalising constants cancel above and below the fraction. Hence, the acceptance probability can be evaluated, unlike in the case of a standard Metropolis-Hastings scheme. In practice, the exchange move proposes to offer the observed data ${\bm x}$ to the auxiliary parameter $\bmlam^{\prime}$ and similarly to offer the auxiliary data ${\bm y}$ the parameter $\bmlam$. 

\section{Applications} \label{sec:application}

\subsection{Artificial Data}  \label{sec:sim_data_Ex1}
We illustrate the proposed algorithm to sample from the posterior distribution of $\bmlam$ using artificial data. \vspace{0.2cm}

{\em Dataset 1} \vspace{0.2cm}

\begin{figure}[H]
\begin{center}
 \begin{tabular}{cc}
  \includegraphics[scale=0.4]{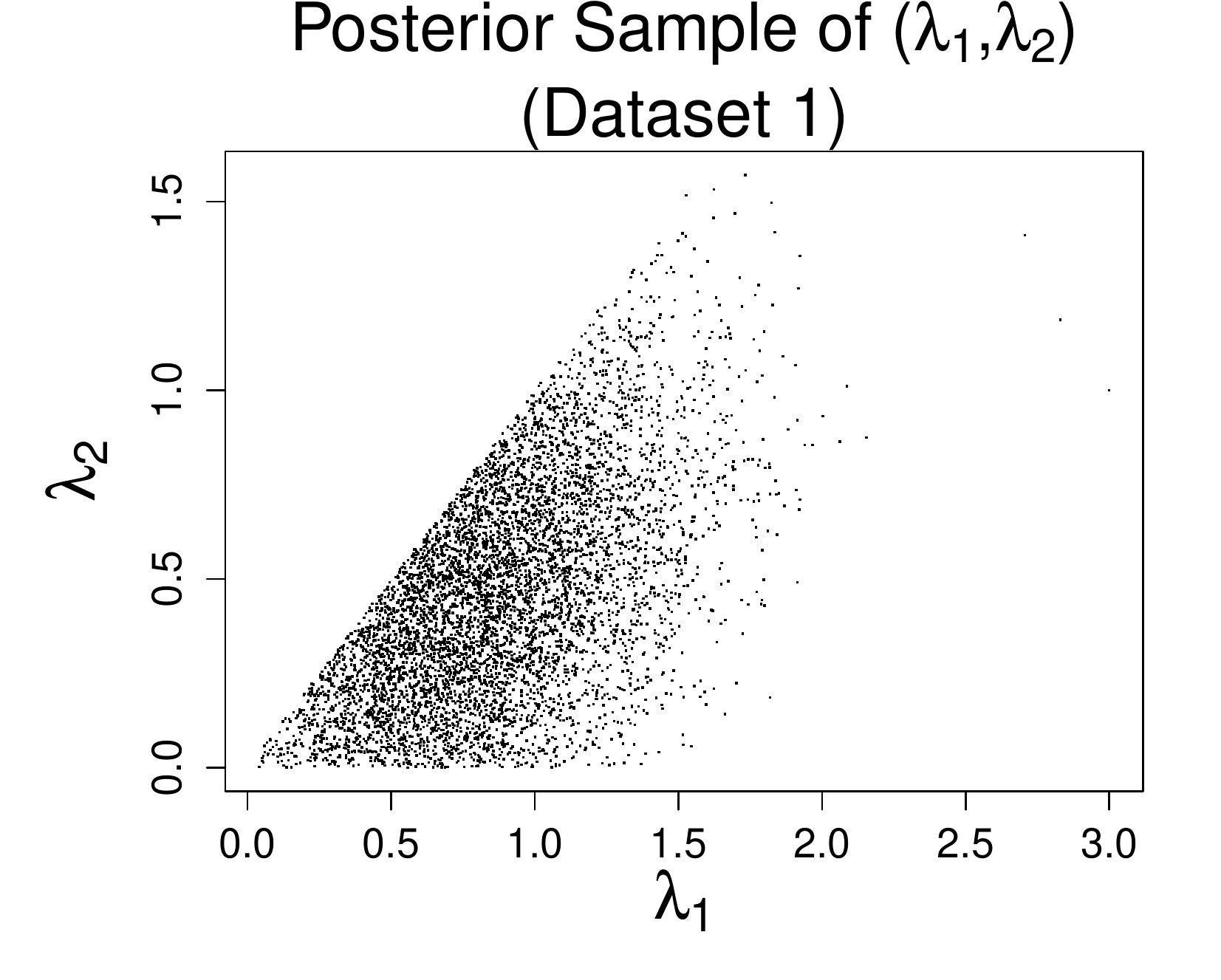} & \includegraphics[scale=0.4]{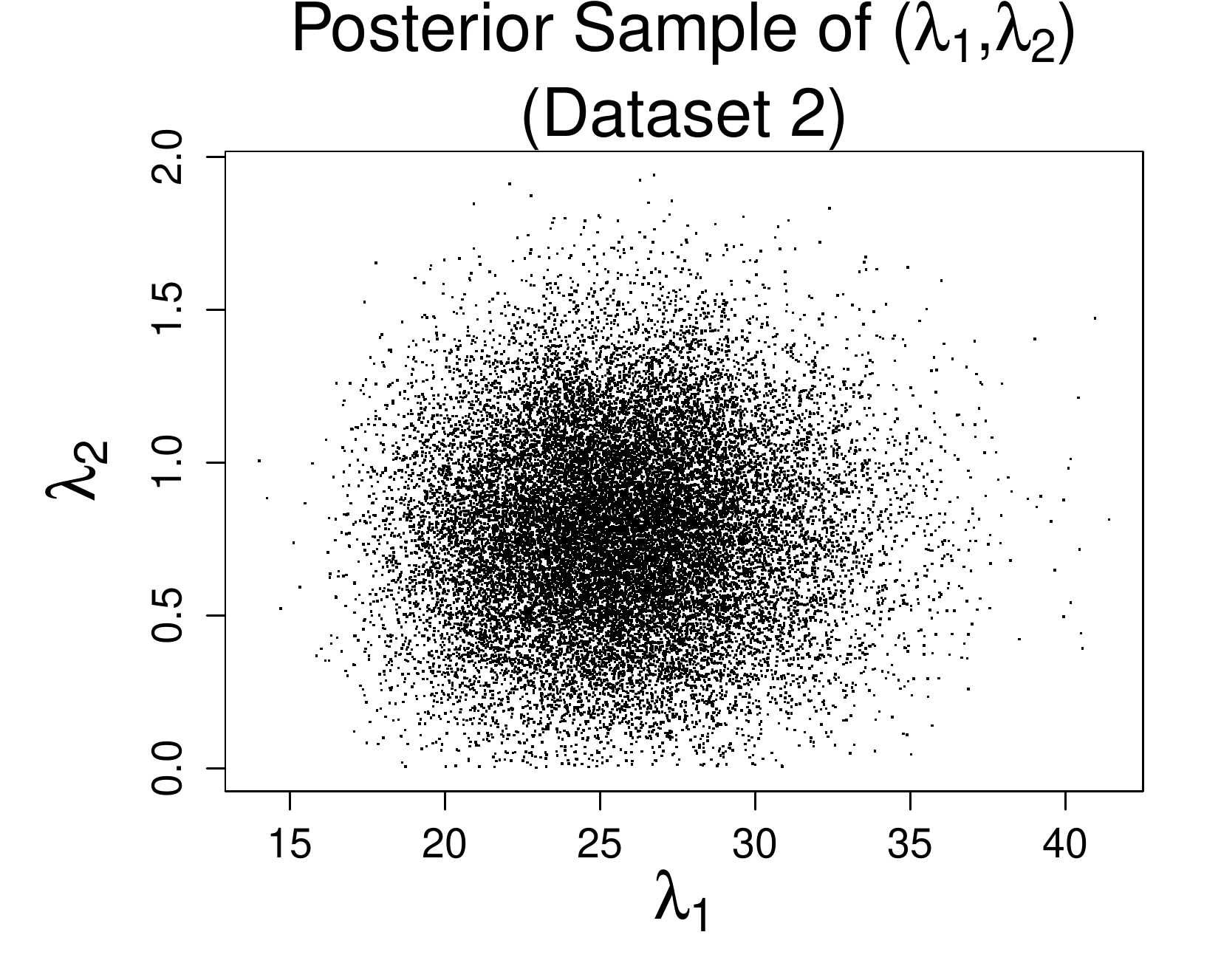}
 \end{tabular}
 \caption{Sample from the joint posterior distribution of $\lambda_1$ and $\lambda_2$ for Dataset 1 (left) and Dataset 2 (right) as described in Section \ref{sec:application}.}\label{fig:sim_data_Ex12_scatter}
 \end{center}
\end{figure}

\begin{figure}[H]
\begin{center}
\begin{tabular}{cc}
 \includegraphics[scale=0.4]{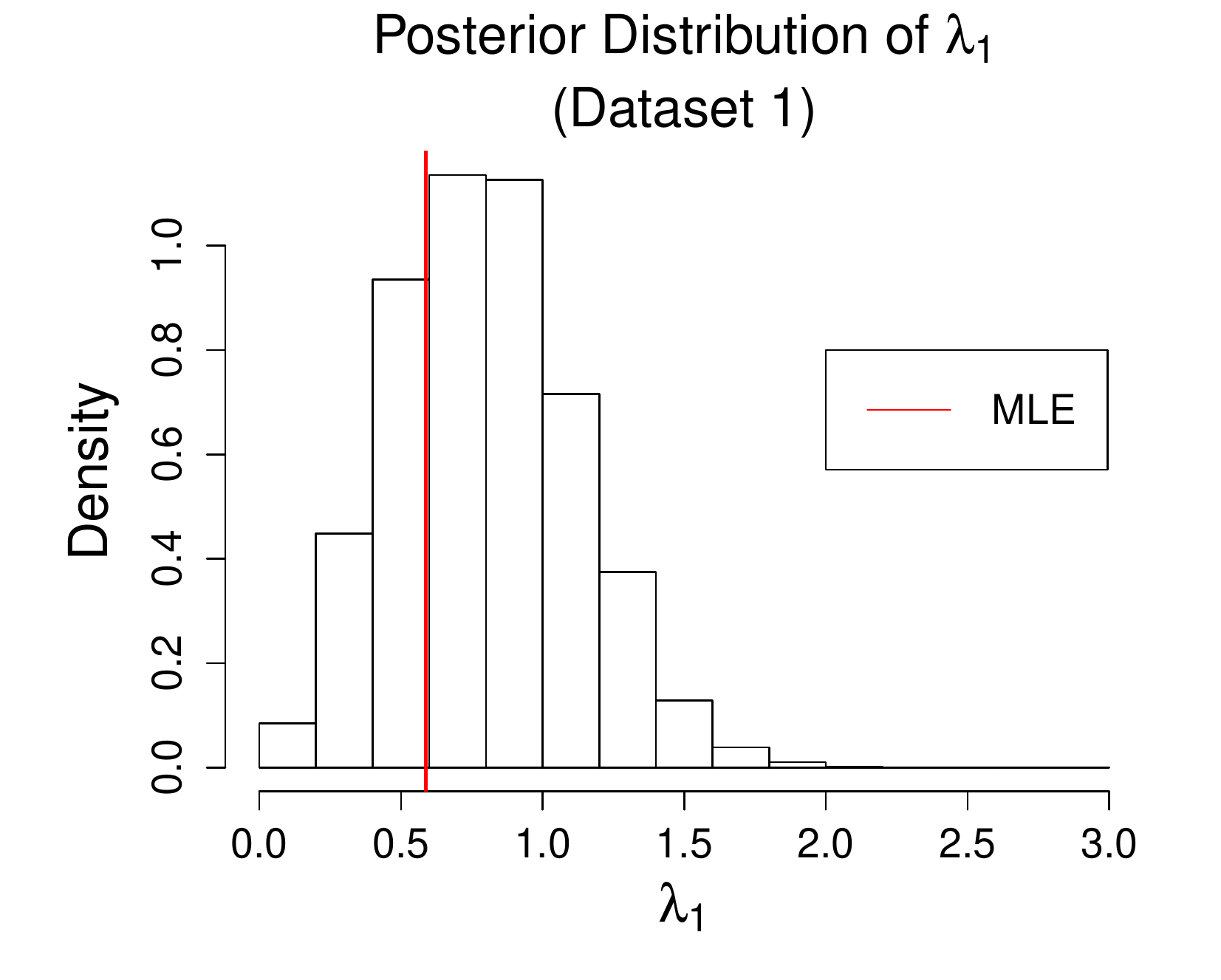} & \includegraphics[scale=0.4]{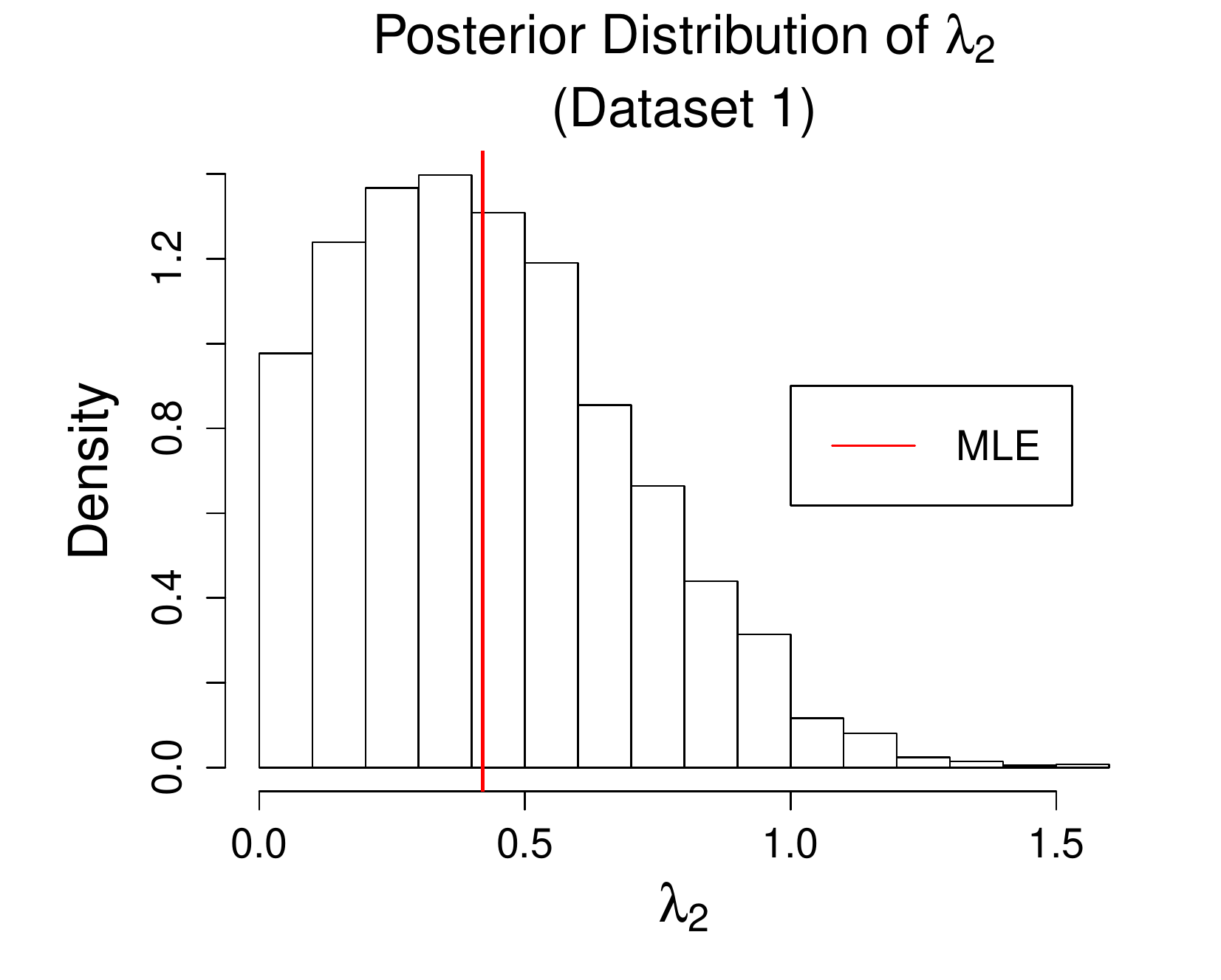}\\
 \includegraphics[scale=0.4]{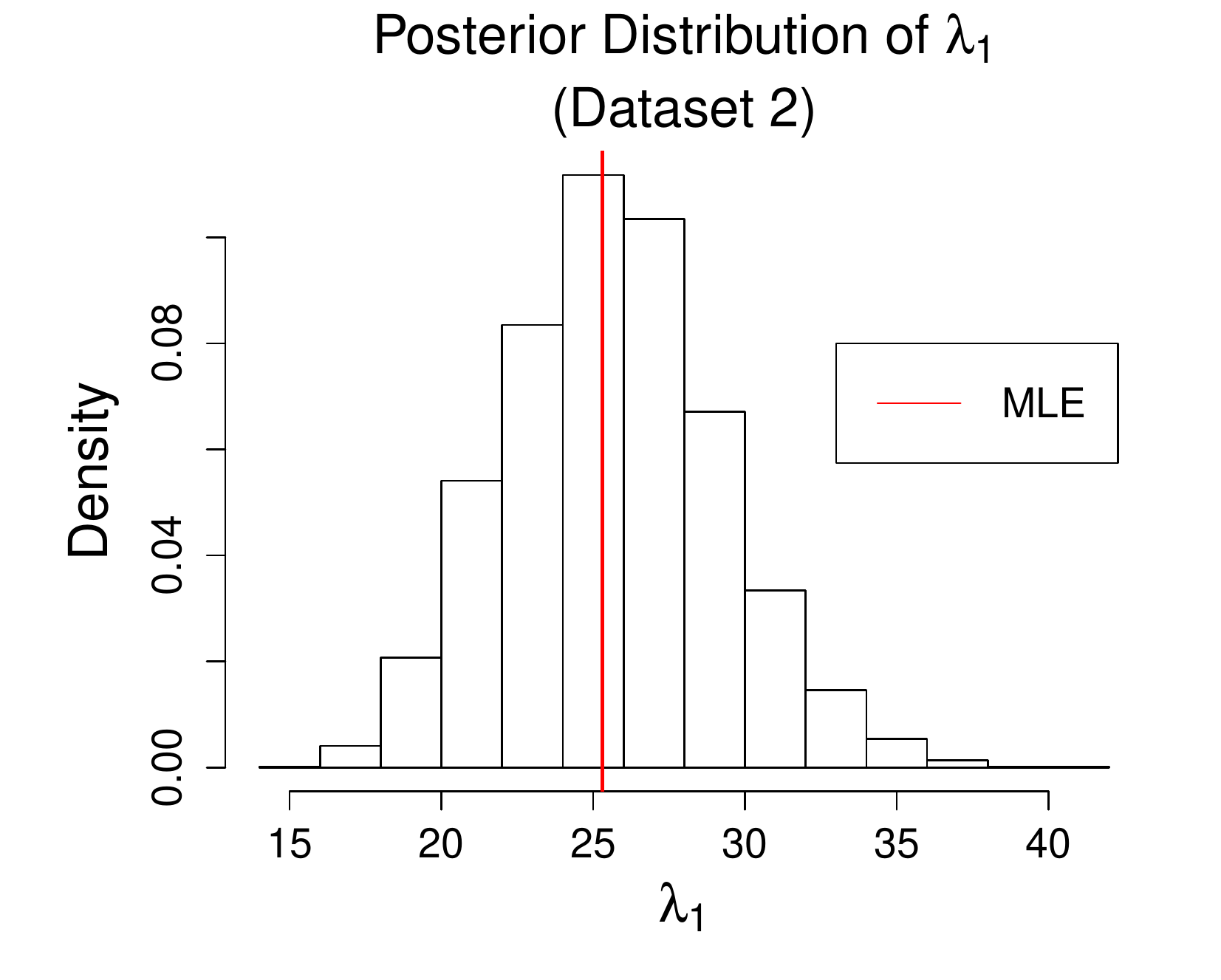} & \includegraphics[scale=0.4]{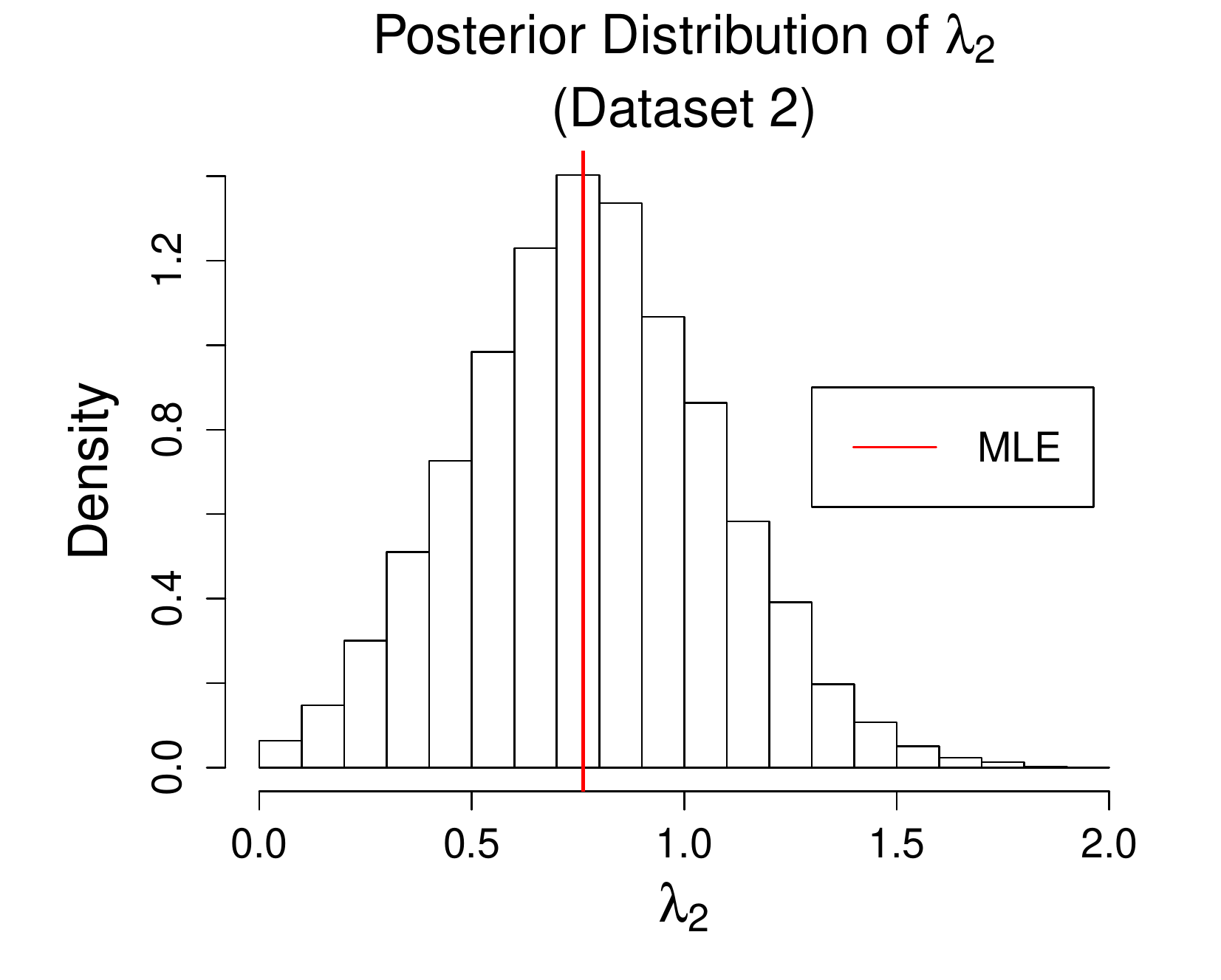}
\end{tabular} \caption{Marginal posterior densities and  ACFs for $\lambda_1$ and $\lambda_2$ for Dataset 1 (top) and Dataset 2 (bottom)	 in  Section \ref{sec:application}.}\label{fig:hist_sim_data_Ex12}
\end{center}
\end{figure}

\begin{figure}[H]
\begin{center}
\begin{tabular}{cc}
 \includegraphics[scale=0.4]{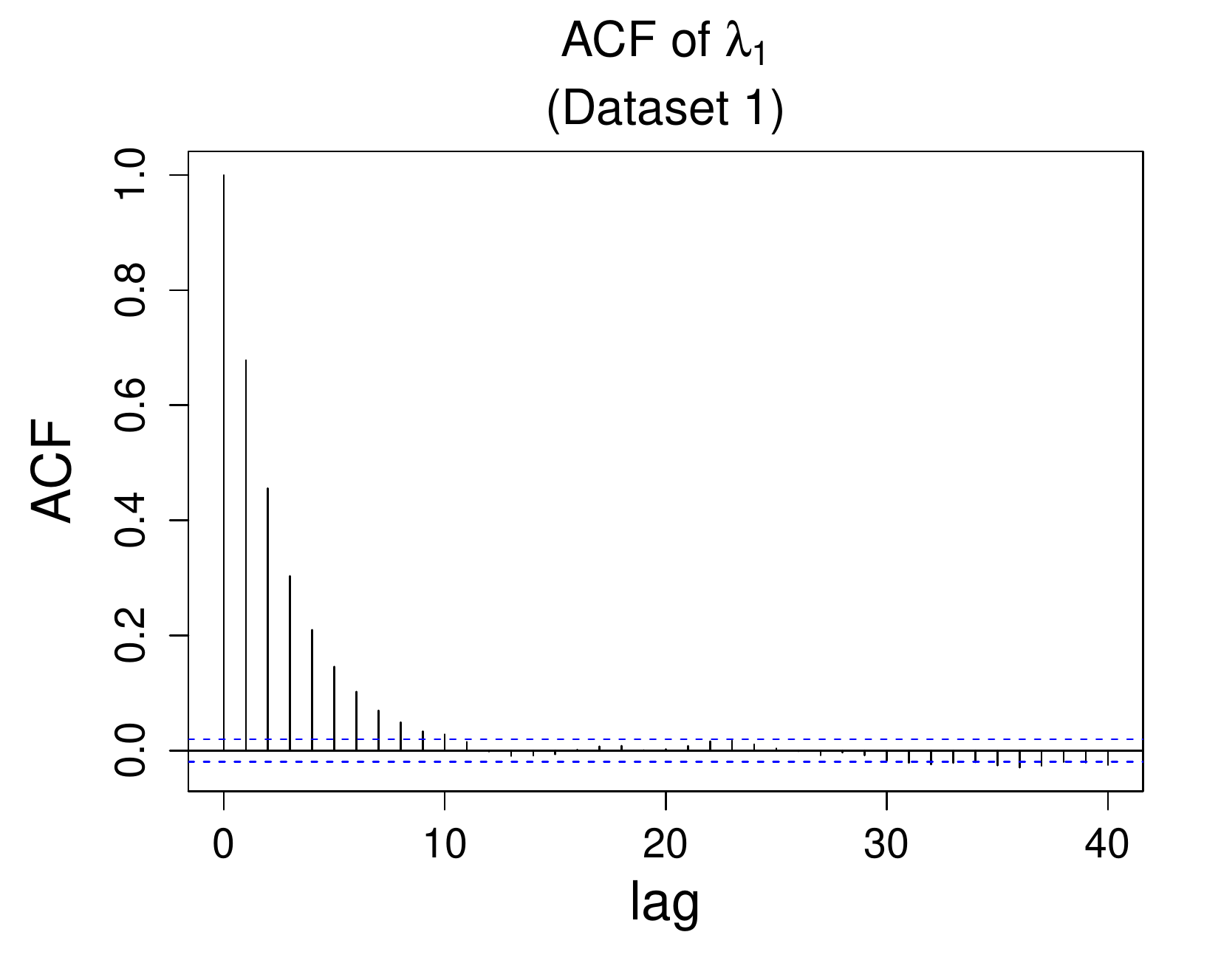} & \includegraphics[scale=0.4]{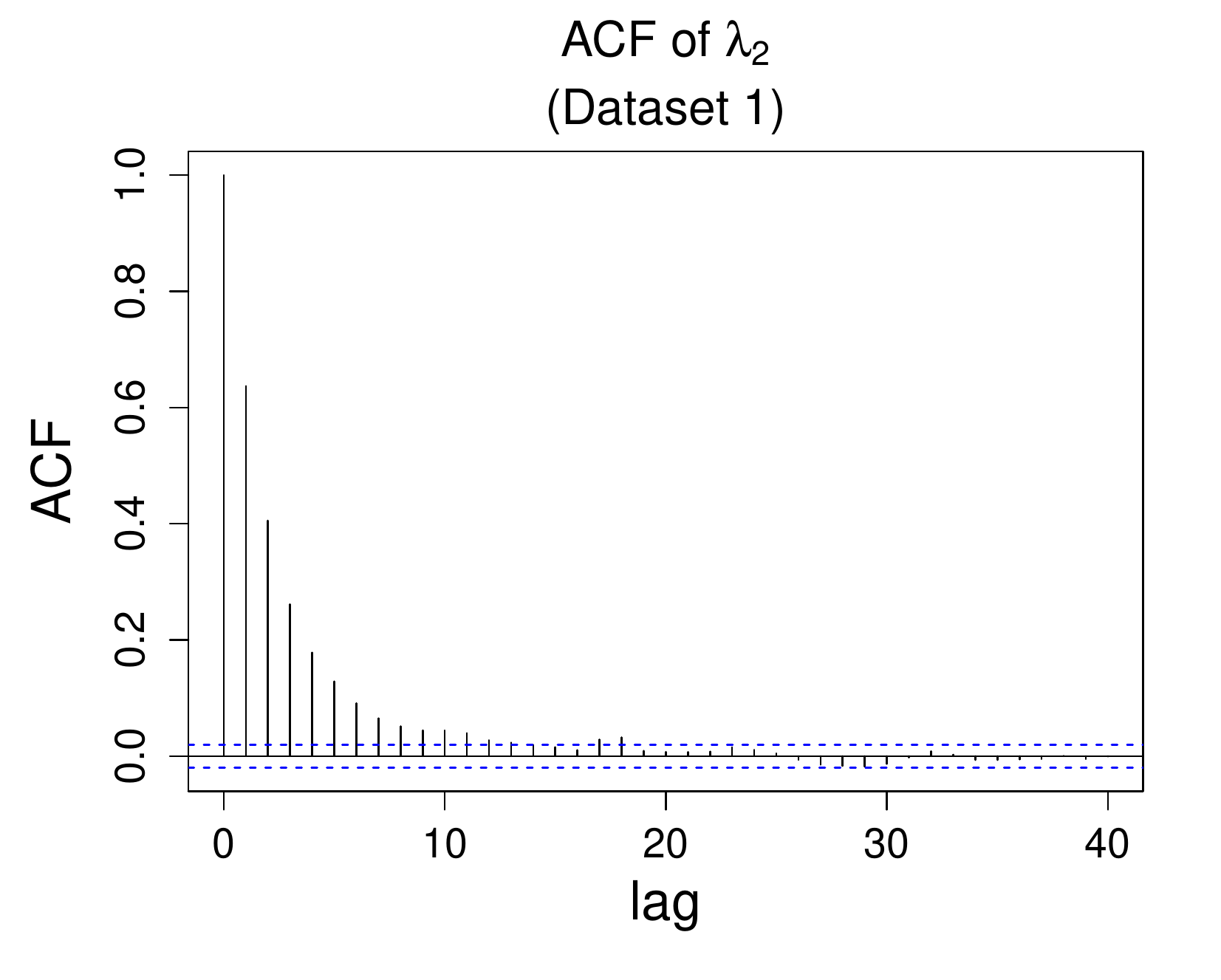} \\ 
 \includegraphics[scale=0.4]{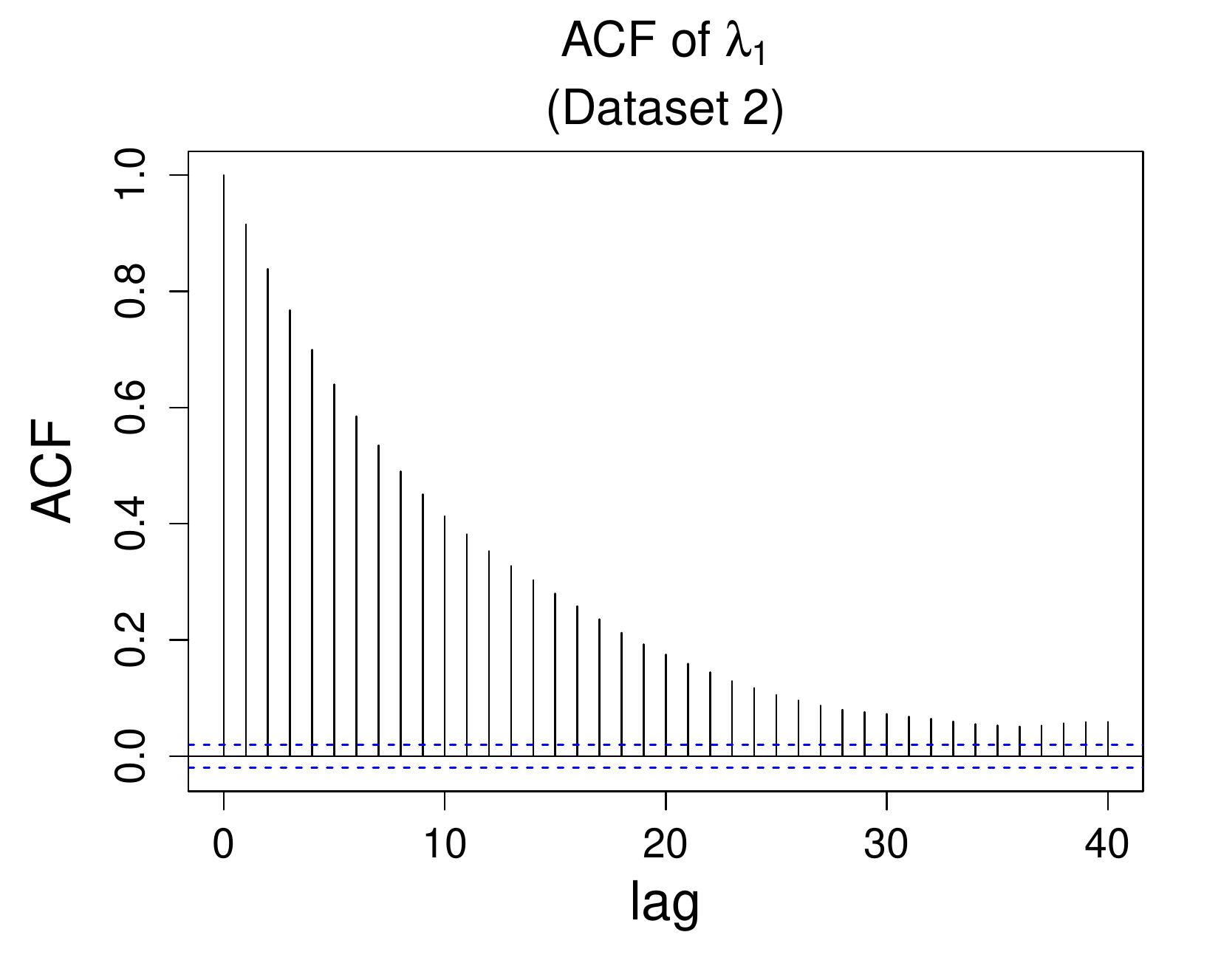} & \includegraphics[scale=0.4]{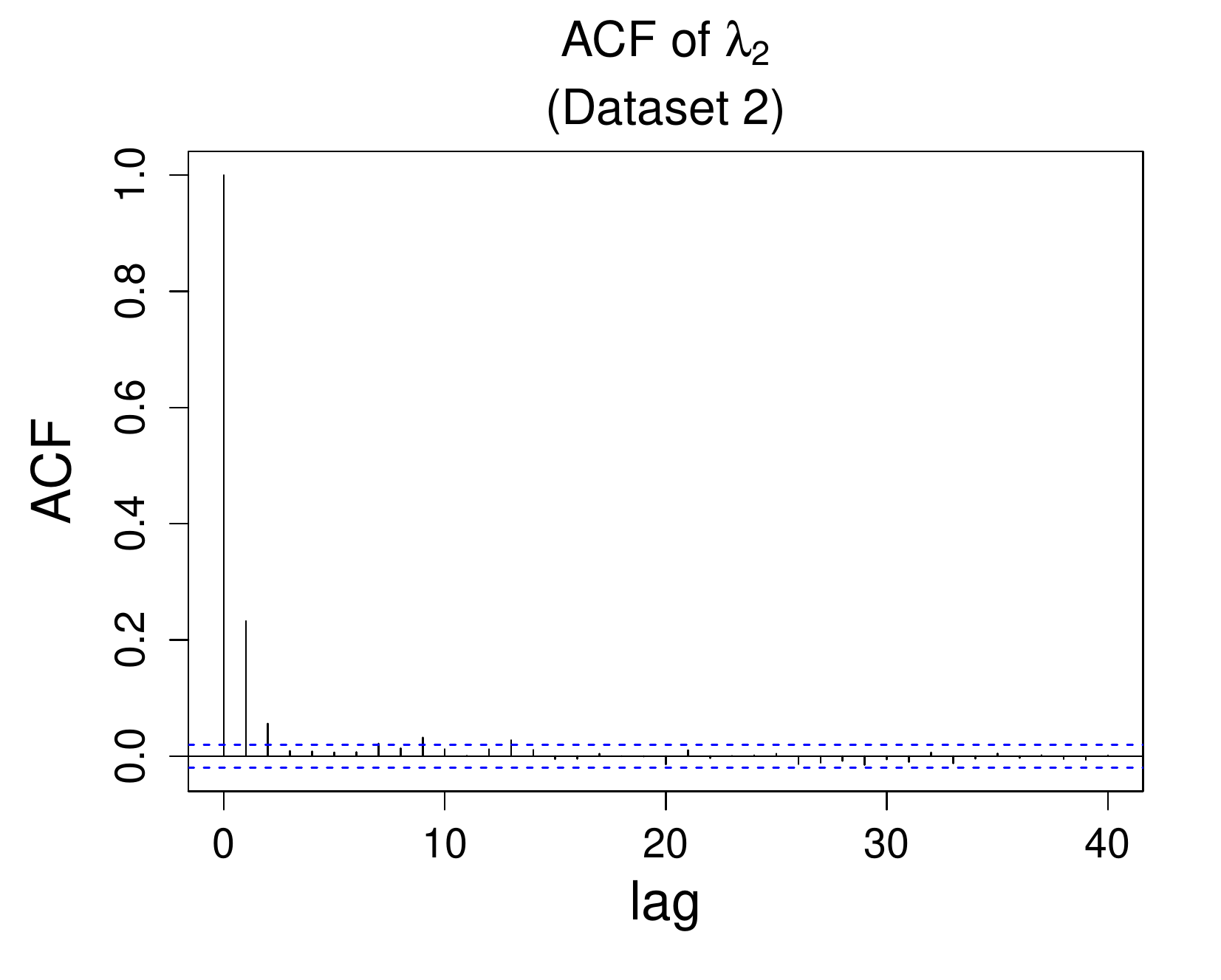} 
\end{tabular} \caption{ACFs for $\lambda_1$ and $\lambda_2$ for Dataset 1 (top) and Dataset 2 (bottom) in  Section \ref{sec:application}.}\label{fig:acf_sim_data_Ex12}
\end{center}
\end{figure}

Consider a sample of $n=100$ unit vectors $(\bmx_1, \ldots, \bmx_{100})$ which result in the pair of sufficient statistics ($\tau_1, \tau_2)=(0.30, 0.32)$. We assign independent Exponential prior distributions with rate $0.01$ (i.e. mean 100) to the parameters of interest $\lambda_1$ and $\lambda_2$ subject to the constraint that $\lambda_1 \geq \lambda_2$; note that we also implicitly assume that $\lambda_1 \geq \lambda_2 \geq\lambda_3 = 0$.  We implemented the algorithm which was described in Section \ref{sec:exchange}. The parameters were updated in blocks by proposing a candidate vector from a bivariate Normal distribution with mean the current values of the parameters and variance-covariance matrix $\sigma I$, where $I$ is the identity matrix and the samples were thinned, keeping every 10th value. Convergence was assessed by visual inspection of the Markov chains  and we found that by using $\sigma=1$ the mixing was good and achieved an acceptance rate between 25\% and 30\%. Figure \ref{fig:sim_data_Ex12_scatter} shows a scatter 
plot of the sample from the 
joint posterior distribution (left panel) whilst the marginal posterior densities  for $\lambda_1$ and $\lambda_2$ are shown in the top row of Figure \ref{fig:hist_sim_data_Ex12}. The autocorrelation function (ACF) plots, shown in the top row of Figure \ref{fig:acf_sim_data_Ex12} reveal good mixing properties of the MCMC algorithm and by (visual inspection) appears to be much better than the algorithm proposed by \citet[Figure 1]{Wal13}. \citet{Mar77} report maximum likelihood estimates of $\hat{\lambda}_1 = 0.588$, $\hat{\lambda}_2 = 0.421$, with which our results broadly agree. Although in principle one can derive (approximate) confidence intervals based on some regularity conditions upon which it can be proved that the MLEs are (asymptotically) Normally distributed, an advantage of our (Bayesian) approach is that it allows quantification of the uncertainty of the parameters of interest in a probabilistic manner. \vspace{0.2cm}

{\em Dataset 2} \vspace{0.2cm}	

We now consider an artificial dataset of 100 vectors which result in the pair of sufficient statistics $(\tau_1, \tau_2)=(0.02, 0.40)$   for which the maximum likelihood estimates are $\hat{\lambda}_1 = 25.31 $, $\hat{\lambda}_2 = 0.762 $ as reported by \citet{Mar77}. We implement the proposed algorithm by assigning the same prior distributions to $\lambda_1$ and $\lambda_2$ as for the Dataset 1.  A scatter plot of a sample from the joint posterior distribution in shown in Figure \ref{fig:sim_data_Ex12_scatter}, showing that our approach gives results which are consistent with the MLEs. This examples shows that our algorithm performs well  even when $\lambda_1 >> \lambda_2$.

\subsection{Earthquake data}\label{sec:earthquake}

As an illustration of an application to real data, we consider an analysis of earthquake data recently analysed by \citet{Arn13}. An earthquake gives rise to three orthogonal axes, and geophysicists are interested in analysing such data in order to compare earthquakes at different locations and/or at different times. An earthquake gives rise to a pair of orthogonal axes, known as the compressional (P) and tensional (T) axes, from which a third axis, known as the null (A) axis is obtained via $A = P \times T$. Each of these quantities are determined only up to sign, and so models for axial data are appropriate. The data can be treated as orthogonal axial $3$-frames in $\mathbb{R}^3$ and analysed accordingly, as in \citet{Arn13}, but we will illustrate our method using the A axes only. In general, an orthogonal axial $r$-frame in $\mathbb{R}^p, \hspace{0.1cm} r \le p $, is an ordered set of $r$ axes, $\{\pm u_1,\ldots,\pm u_r\}$, where $u_1,\ldots,u_r$ are orthonormal vectors in $\mathbb{R}^p$ \citep{Arn13}. 
The more familiar case of data on the sphere $\mathcal{S}^2$ is the special case corresponding to $p=3,r=1$, which is the case we consider here.

The data consist of three clusters of observations relating to earthquakes in New Zealand. The first two clusters each consist of $50$ observations near Christchurch which took place before and after a large earthquake on 22 February 2011, and we will label these two clusters CCA and CCB respectively. For these two clusters, the P and T axes are quite highly concentrated in the horizontal plane, and as a result the majority of the A axes are concentrated about the vertical axis. It is of interest to geophysicists to establish whether there is a difference between the pattern of earthquakes before and after the large earthquake. The third cluster is a more diverse set of $32$ observations obtained from earthquakes in the north of New Zealand's South Island, and we will label this cluster SI. We will illustrate our method by fitting Bingham models to the A axes from each of the individual clusters and considering the posterior distributions of the Bingham parameters. We will denote the parameters from the CCA, 
CCB and SI models as $\lambda_i^A$, $\lambda_i^B$ and $\lambda_i^S$ respectively, $i=1,2$. 

\begin{figure}[H]
\begin{center}
\begin{tabular}{ccc}
 \includegraphics[scale=0.3]{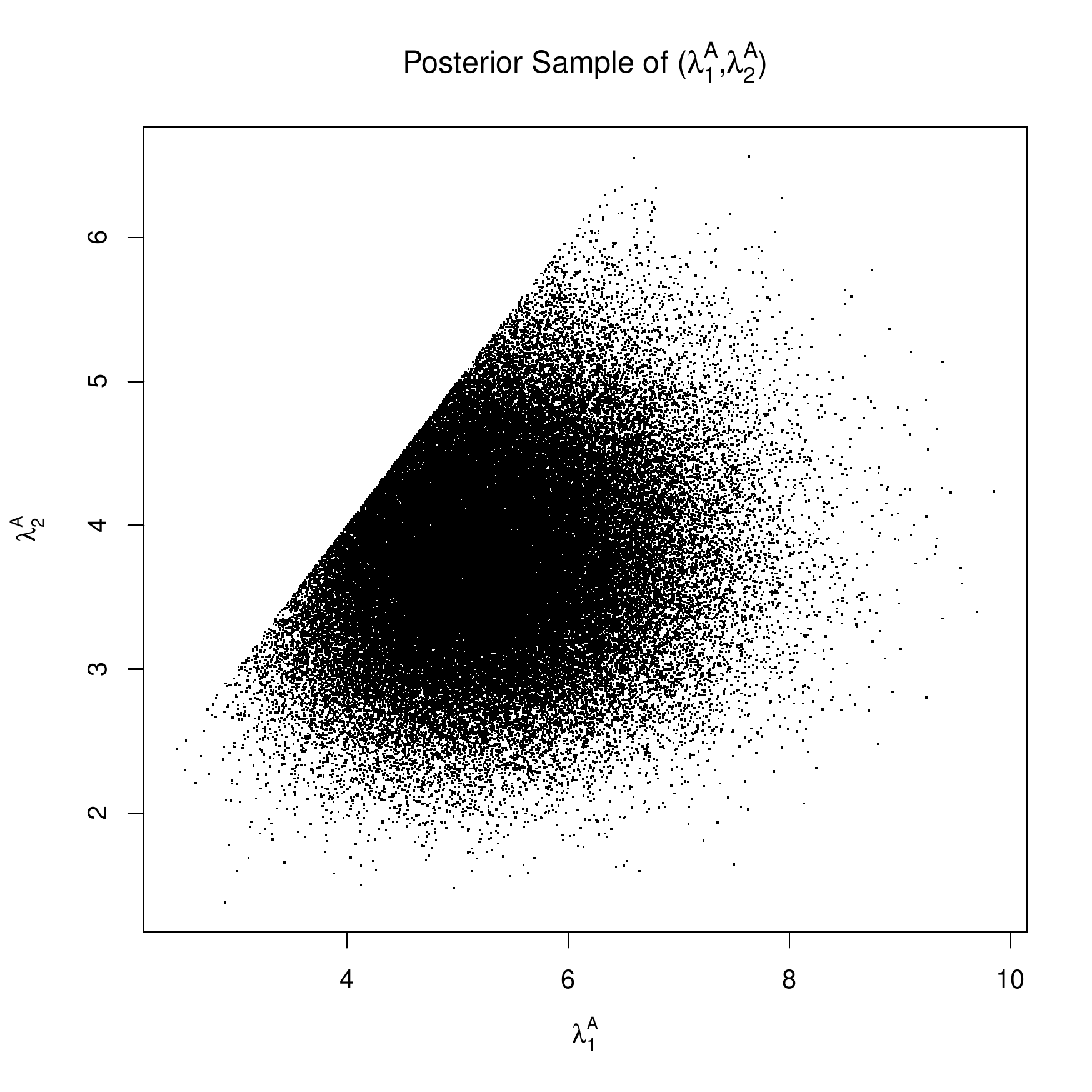} & \includegraphics[scale=0.3]{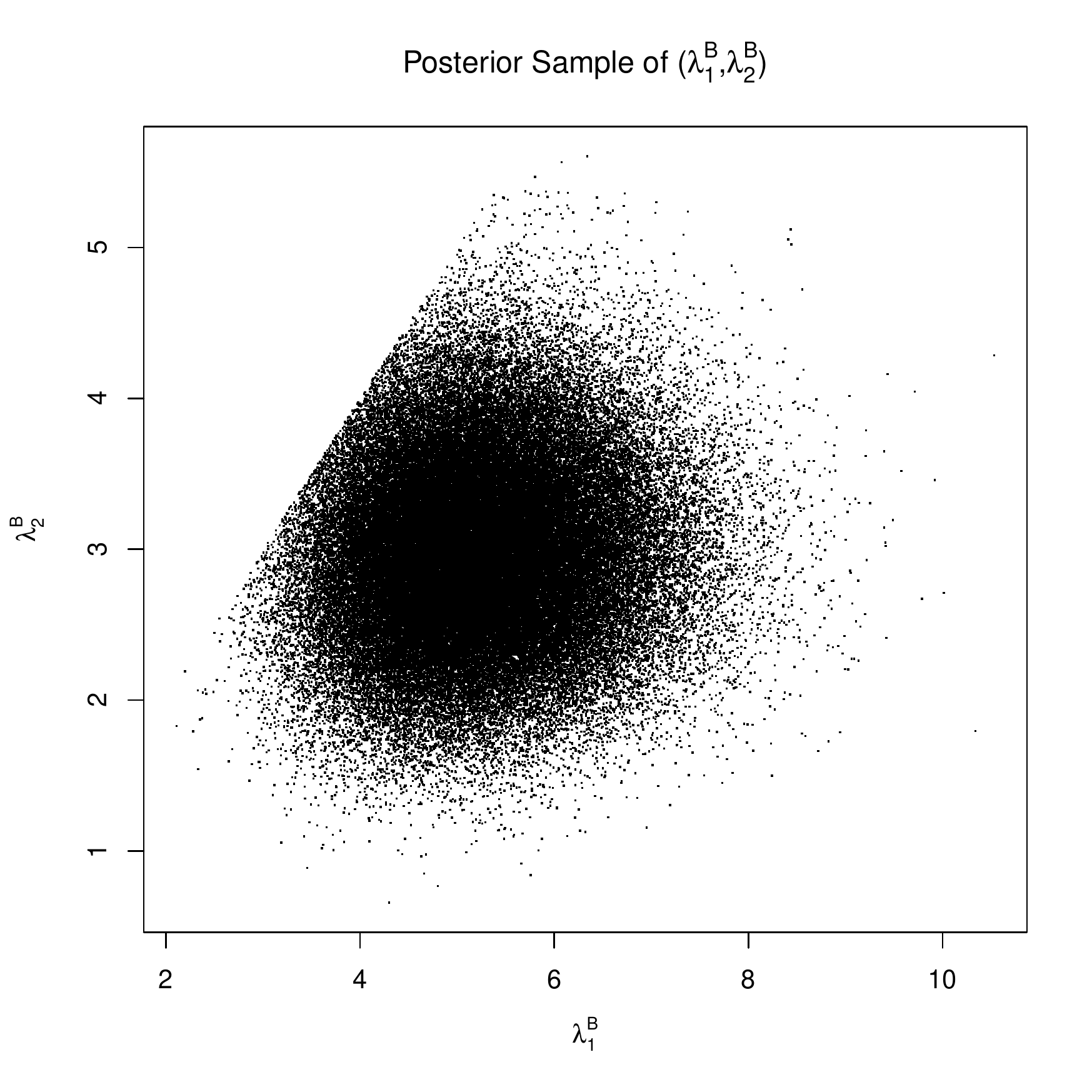} & \includegraphics[scale=0.3]{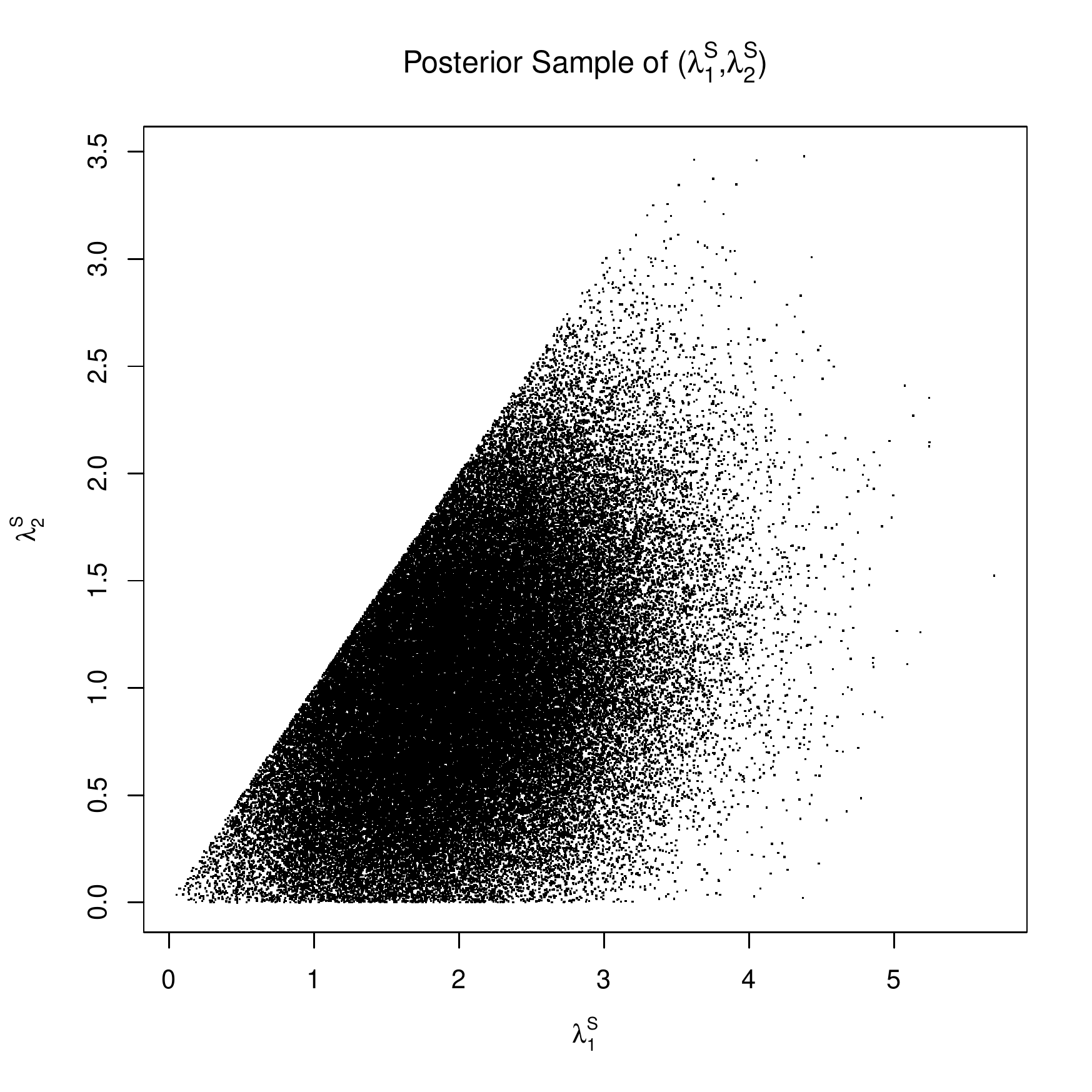}
\end{tabular} 
\caption{Posterior samples for differences in $\lambda_1$ and $\lambda_2$ for the two sets of Christchurch data (left) and South Island and Christchurch data A (right). This shows a clear difference between the South Island and Christchurch data, but suggests no difference between the two sets of Christchurch data.}\label{fig:Earthquake_Scatter}
\end{center}
\end{figure}

The observations for the two clusters of observations near Christchurch yield sample data of $(\tau_1^A, \tau_2^A) =(0.1152360,0.1571938)$ for CCA  and $(\tau_1^B, \tau_2^B)(0.1127693,0.1987671)$ for CCB. The data for the South Island observations are  $(\tau_1^S, \tau_2^S)=(0.2288201,0.3035098)$. We fit each dataset separately by implementing the proposed algorithm.Exponential prior distributions to all parameters of interest (mean 100) were assigned, subject to the constraint that $\lambda_1^j \geq \lambda_2^j$ for $j=A, B, S$. Scatter plots from the joint posterior distributions of the parameters from each individual analysis are shown in Figure \ref{fig:Earthquake_Scatter}. The plots for CCA and CCB look fairly similar, although $\lambda_2$ is a little lower for the CCB cluster. The plot for SI cluster suggests that these data are somewhat different.             

\begin{figure}[H]
\begin{center}
\begin{tabular}{cc}
 \includegraphics[scale=0.5]{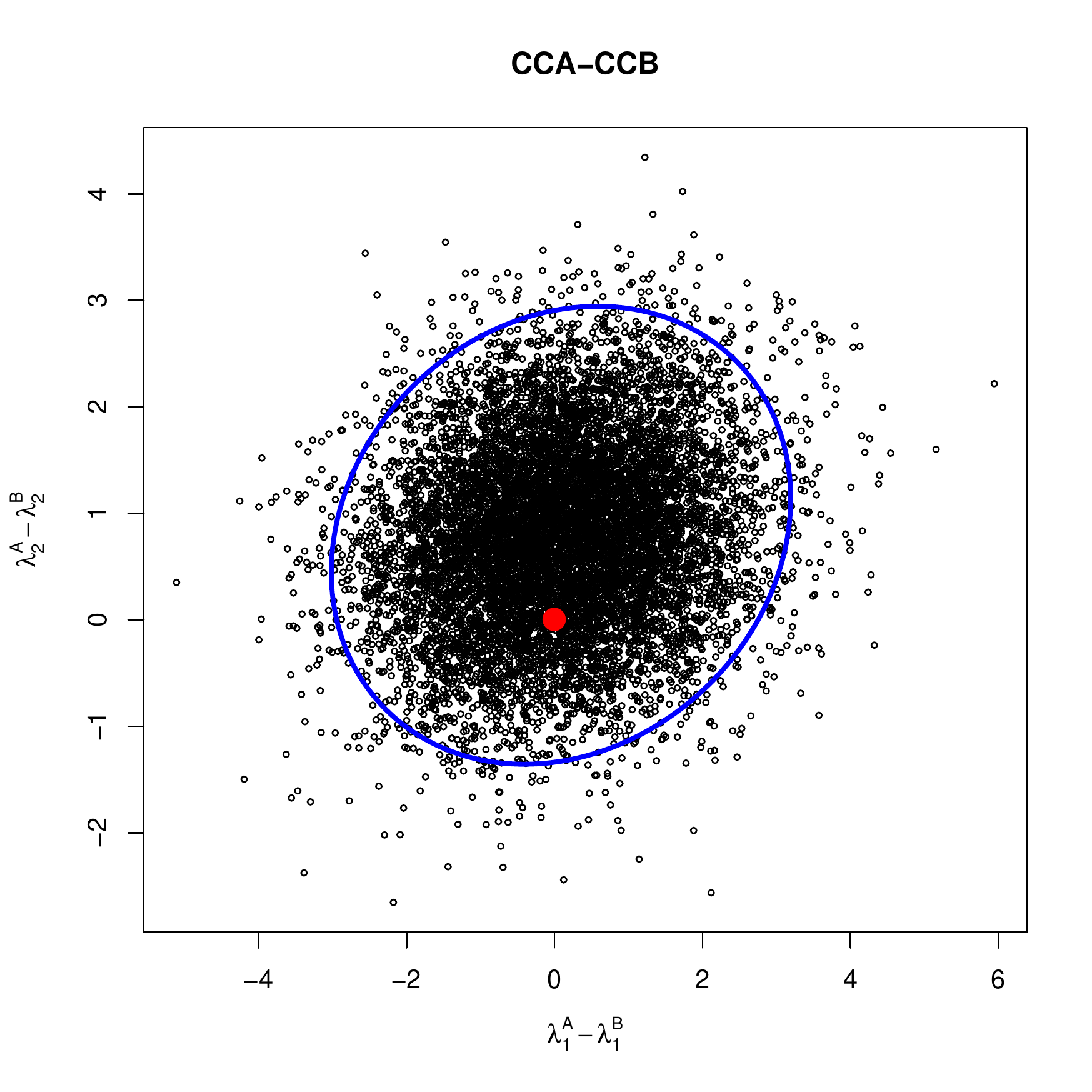} & \includegraphics[scale=0.5]{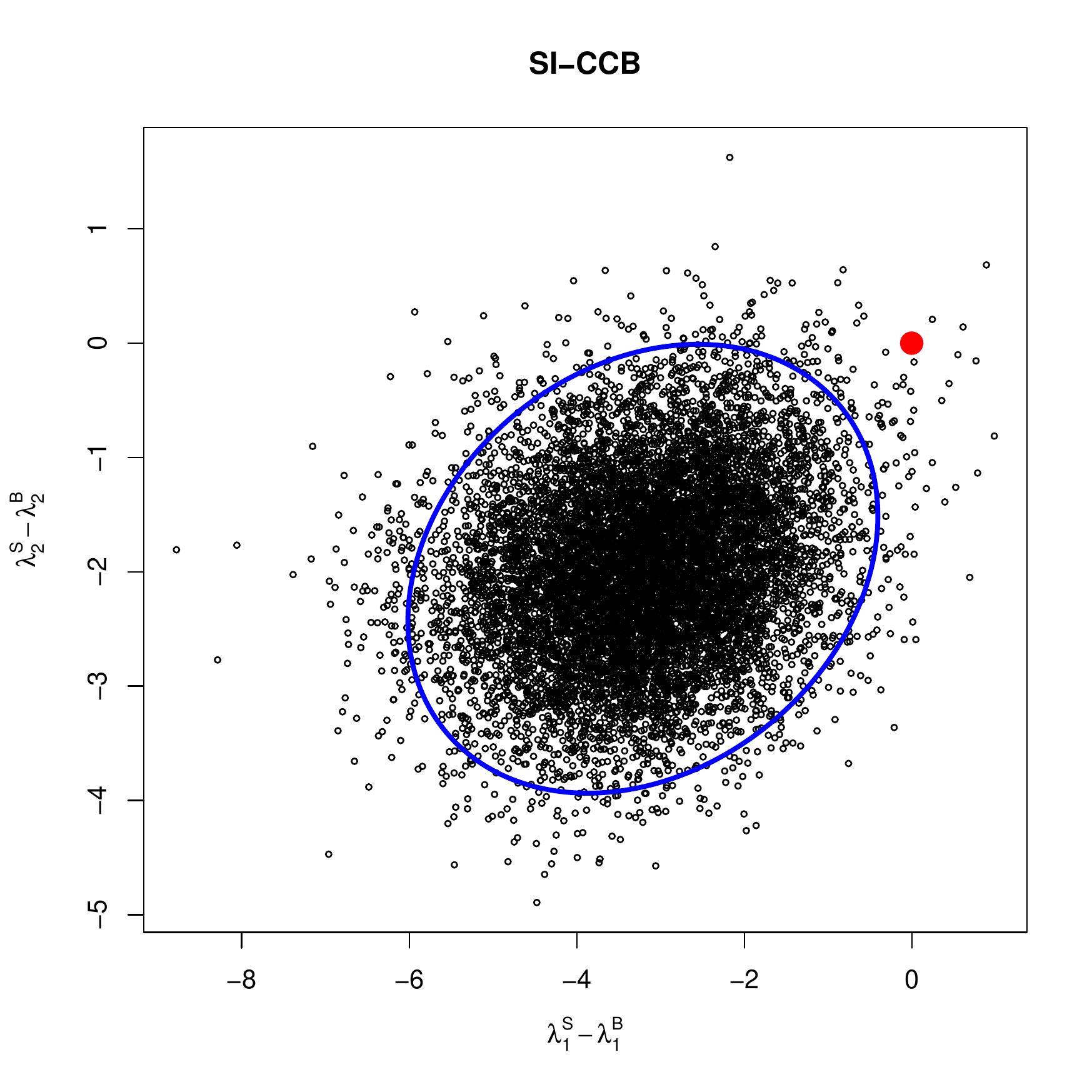}
\end{tabular} 
\caption{Posterior samples for differences in $\lambda_1$ and $\lambda_2$ for the two sets of Christchurch data (left) and South Island and Christchurch data A (right). This shows a clear difference between the South Island and Christchurch data, but suggests no difference between the two sets of Christchurch data.}\label{fig:Earthquake_test}
\end{center}
\end{figure}

To establish more formally if there is any evidence of a difference between the two Christchurch clusters, we consider the bivariate quantity $(\lambda_1^A - \lambda_1^B,\lambda_2^A - \lambda_2^B).$ If there is no difference between the two clusters, then this quantity should be $(0,0)$. In Figure \ref{fig:Earthquake_test} (left panel), we show the posterior sample of this quantity, and a 95\% probability region obtained by fitting a bivariate normal distribution with parameters estimated from this sample. The origin is contained comfortably within this region, suggesting there is no real evidence for a difference between the two clusters.  \citet{Arn13}  obtained a $p$-value of $0.890$ from a test of equality for the two populations based on treating the data as full axial frames, and our analysis on the A axes alone agrees with this. 

The right panel of Figure \ref{fig:Earthquake_test} shows a similar plot for the quantity $(\lambda_1^A - \lambda_1^S,\lambda_2^A - \lambda_2^S).$ Here, the origin lies outside the 95\% probability region, suggesting a difference between the first Christchurch cluster and the South Island cluster. \citet{Arn13} give a $p$-value of less than 0.001 for equality of the two populations, so again our analysis on the A axes agrees with this. 

\section{Discussion} \label{sec:concl}
There is a growing area of applications that require inference over doubly intractable distributions including directional statistics, social networks \citep{CaimFrie11}, latent Markov random fields \citep{Ever12}, and large--scale spatial statistics \citep{AunSimEid12} to name but a few. Most conventional inferential methods for such problems relied on approximating the normalising constant and embedded the latter into a standard MCMC algorithm (e.g. Metropolis-Hastings). Such approaches not only are only approximate in the sense that the target distribution is an approximation to the true posterior distribution of interest, but they can also suffer from being very computationally intensive. It is only until fairly recently that algorithms which avoid the need of approximating/evaluating the normalising constant became available; see \citet{Mol_etal06, MurGhahMac06, Wal11, Gir_etal13}.

In this paper we were concerned with exact Bayesian inference for the Bingham distribution which has been a difficult task so far. We proposed an MCMC algorithm which allows us to draw samples from the posterior distribution of interest without having to approximate this constant. We have shown that the MCMC scheme is i) fairly straightforward to implement, ii) mixes very well in a relatively short number of sweeps and iii) does not require the specification of good guesses of the unknown parameters. We have applied our method to both real and simulated data, and showed that the results agree with maximum likelihood estimates for the parameters. However, we believe that a fully Bayesian approach has the benefit of providing an honest assessment of the uncertainty of the parameter estimates and allows exploration of any non-linear correlations between the parameters of interest. In comparison to the approach recently proposed by \citet{Wal13} (which also avoids approximating the normalising constant) we argue 
that our algorithm is easier to implement, runs faster and the Markov chains appear to mix better.

%\subsection{Computational considerations}
In terms of the further computational aspects, our algorithm is not computationally intensive and this is particularly true for the number of dimensions that are commonly met in practice (e.g. $q=3$). For all the results presented here, we ran our MCMC chains for $10^6$ iterations for each of the simulated and real data examples, which we found to be sufficient for good mixing in all cases. Our method was implemented in C++ and each example took between $20$ and $30$ seconds on a desktop PC with 3.1GHz processor\footnote{Our code is available upon request.}; note, that is considerably faster than the algorithm proposed by \citet{Wal13} in which ``running $10^5$ iterations takes a matter of minutes on a standard laptop''. In general the time taken for our proposed algorithm will depend on the number of auxiliary data points $n$ that need to be simulated, as well as the efficiency of the underlying rejection algorithm for the particular parameter values at each iteration. In addition, the efficiency of the 
rejection algorithm is likely to deteriorate as the dimension $q$ increases, but we found it to be very efficient for all our examples and it is reasonably efficient for at least a moderate number of dimensions according to simulations by \citet{Gan12}. 

Statistical inference, in general, is not limited to parameter estimation. Therefore, a possible direction for future research within this context is to develop methodology to enable calculation of the model evidence (marginal likelihood). This quantity is vital in Bayesian model choice and knowledge of it will allow a formal comparison between competing models for a given dataset such as the application presented in Section \ref{sec:earthquake}.  

\section*{Acknowledgements}
The authors are most grateful to Richard Arnold and Peter Jupp for providing the earthquake data and John Kent for providing a Fortran program to compute moments of the Bingham distribution. Finally, we would like to thank Ian Dryden for commenting on an earlier draft of this manuscript.

\bibliographystyle{dcu}
\bibliography{refs}

\end{document}